\def\be{\begin{equation}}
\def\ee{\end{equation}}
\def\bea{\begin{eqnarray}}
\def\eea{\end{eqnarray}}
\def\bse{\begin{subequations}}
\def\ese{\end{subequations}}

\documentclass[prl,twocolumn,showpacs,preprintnumbers,amsmath,amssymb]{revtex4}
\usepackage{graphicx}
\usepackage{dcolumn}
\usepackage{bm}
\usepackage{wasysym}
\begin{document}
\title{Emergence of Artificial Photons in an Optical Lattice\\
\vskip 1mm
}
\author{Sumanta Tewari$^{1}$, V. W. Scarola$^{1}$, T. Senthil$^{2,3}$ and
        S. Das Sarma$^{1}$}
\affiliation{$^{1}$Condensed Matter Theory Center, Department of
Physics, University of Maryland, College Park, MD 20742\\
             $^{2}$Center for Condensed Matter Theory, Indian Institute
of Science, Bangalore 560012, India\\
         $^{3}$Department of
Physics, Massachusetts Institute of Technology, Cambridge,
Massachusetts 02139}

\date{\today
}
\begin{abstract}
We establish the theoretical feasibility of direct analog simulation
of the compact $U(1)$ lattice gauge theories in optical lattices
with dipolar bosons. We discuss the realizability of the topological
Coulomb phase in extended Bose-Hubbard models in several optical
lattice geometries. We predict the testable signatures of this
emergent phase in noise correlation measurements, thus suggesting
the possible emergence of artificial light in optical lattices.
%
\end{abstract}


\maketitle

{\em Introduction.} Cold atomic gases in optical lattices
\cite{Phillips} have provided unprecedented flexibility in designing
and studying coherent and correlated condensed matter systems. To
date, experiments on strongly correlated, bosonic lattice models
have realized the on-site Hubbard-type interaction $U$, comparable
to a nearest-neighbor hopping $t$. Consequently, Several strongly
correlated bosonic phases with short range interaction have been
studied \cite{Jaksch-Bruder,Greiner,Paredes,Porto}. However, with
the recent observation of Bose-condensation of Chromium
\cite{Griesmaier}, which has a magnetic dipolar interaction, a
sizable nearest-neighbor interaction $V$, albeit anisotropic, is now
 within experimental reach.  Moreover, since $U$ can be tuned using
 Feshbach resonances \cite{Jaksch-Zoller}, $U/V$
can, in principle, be made to vary over a wide range. Furthermore,
exciting developments \cite{molecules} in cooling polar molecules
offer the possibility of strong and tunable electrical dipole
moments.

Recently, it has been proposed \cite{Buchler}, that a four-site
`ring-exchange' interaction can also be implemented
 allowing the simulation of a $U(1)$-lattice gauge theory, which has
various exotic topological phases, among them a 3D $U(1)$ Coulomb
phase with an emergent massless photon mode. In this Letter, we show
that by simply implementing a Hubbard model with an additional
strong nearest-neighbor interaction on some special lattices in both
2D and 3D one can efficiently simulate a $U(1)$-lattice gauge
theory. The dipole interaction, $V_{dd}(R)
=d^2[1-3\cos^2(\phi)]/R^3$, where $d$ characterizes the dipole
moment and $R$ is the inter-dipole separation, is, in general,
anisotropic and depends on the angle $\phi$ between the vectors
defining the parallel dipole orientation and the bond between the
two dipoles. This interaction needs to be made isotropic for our
purpose. This is easily done in 2D by simply aligning the dipoles
perpendicular to the plane. In 3D, this is possible only on some
lattice and we take the pyrochlore lattice where this can be
achieved. Once the gauge theory is simulated, the Coulomb phase can
be accessed by the appropriate tuning of interaction parameters. The
existence of a gauge theory automatically implies that the
conventional insulating phases \cite{Goral} are not the only phases
possible.

We predict and discuss the observable
signatures of the Coulomb phase, specifically, how the emergent photon
mode, giving rise to artificial electrodynamics \cite{Senthil1,Senthil2,Wen},
can be detected in noise correlation measurements
\cite{Eugene,Foelling,Spielman}.  We note, in passing, that other
fractionalized insulators of the gauge theory, where the elementary
excitations carry fractional boson `charge' (boson `charge' in the
present context of neutral bosons means boson number), can also be
accessed in other regions of the parameter space. These, and
experimental implications thereof, will be discussed in a future
work \cite{future_pub}.

\begin{figure}
\includegraphics[scale=0.35]{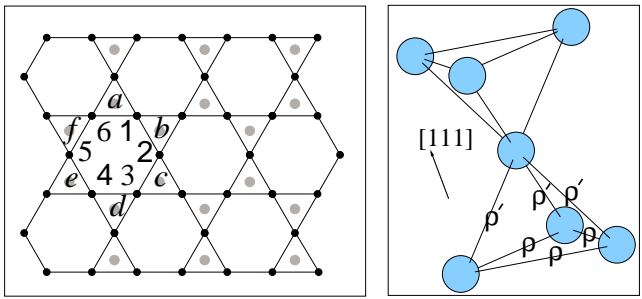}
\caption{ In the left figure dipolar bosons sit on the sites of the
2D Kagom\'{e} lattice, the black circles labeled 1-6.
The hexagonal
dual lattice, the lattice formed by the centers of the
corner-sharing triangles (gray sites), is labeled $a-f$.  In the
right figure, for simplicity, we show only two corner-sharing
tetrahedra of the pyrochlore lattice, with sites shown as spheres.
}
\label{fig:lattice}
\end{figure}

{\em Dipolar Bosons in Optical Lattices.}  We study the Coulomb
insulating phase of dipolar bosons in optical lattices, specifically
the Kagom\'{e} (2D corner-sharing triangles) and pyrochlore (3D
corner-sharing tetrahedra) lattices, see Fig.~\ref{fig:lattice}.
Optical lattices formed from the intensity extrema of standing wave
lasers can be used to generate a variety of lattices with
non-trivial primitive unit cells using the superlattice technique
\cite{Santos}, originally proposed to generate two dimensional
lattices.  Using this technique one can, in principle, generate a
well defined set of intensity maxima defining the Kagom\'{e} and the
regular triangular lattices through angular interference of several
beams with the same wavelength.  Remarkably, the potentials defining
these two lattices only differ by a phase. By generalizing this
technique to three dimensions one can create a three-dimensional
pyrochlore lattice by alternately stacking the triangular and
Kagom\'{e} lattices. Consider the following intensity pattern
generated by counter-propagating, red-detuned lasers: $
F_i(\vec{r})=[\cos(\vec{k}_i\cdot\vec{r}+\Phi_i)+2\cos(\vec{k}_i\cdot\vec{r}/3+\Phi_i/3)]^2.
$ With polarization (or frequency) mismatches, intensities can be
added to generate the following three-dimensional potential with
maxima defining a pyrochlore lattice: $ V_p(\vec{r})\propto
\sum_{i=1}^{5}B_i F_i(\vec{r}). $ The relative intensities and
wavevectors are given by: $ B_{i\neq 4}=1 $, $B_4=2B_1$,
$\vec{k}_1=\bar{k}(1,0,-5/\gamma)$,
$\vec{k}_{2,3}=\bar{k}(1,\pm\sqrt{3}/2,-1/\gamma)$, and
$\vec{k}_{5}=\vec{k}_{4}=\bar{k}(0,0,3/\gamma)$. Here we have
defined $\gamma\equiv 4[1-\rho^2/(3\rho'^2)]^{1/2}$. The relative
phases of the standing waves, $\Phi_{i=2-4}=0$ and
$\Phi_1=\Phi_5=-3\pi/2$, must also be fixed. To generate $V_p$, we
require phase matching between beams with varying spatial
periodicities.  The fact that the beams must also be tuned,
energetically, to internal states of constituent bosons (which
normally fixes the periodicity) poses an experimental challenge.
However, recent experiments have in fact demonstrated multiple
spatial periodicities through polarization-selective, angled
interference \cite{Porto} and Doppler-sensitive multi-photon
transitions \cite{Ritt}, separately. A combination of both
techniques could offer a versatile system capable of realizing the
standing wave arrangement used to produce $V_p$.  Further details
discussing the parameters and laser geometry for these lattices will
be discussed elsewhere \cite{future_pub}.

The interaction between dipolar bosons confined to optical lattices
arises from two terms: the short range, $s$-wave, interaction,
$V_s(R) = g\delta^{(3)}(R)$, and the long range dipolar interaction
$V_{dd}(R)$.  In what follows we place the system in a uniform external field,
pointing along the stacking direction (the $[111]$ direction for
the pyrochlore lattice) to orient the
dipoles perpendicular to the basal plane.  We also
assume that the interaction strength $g$ is tunable through Feshbach
resonances, allowing tunability of the ratio
$V_s/V_{dd}$.  To make the dipolar
interaction, $V_{dd}(R)$, isotropic
we need a pyrochlore lattice where the tetrahedra are shrunk in the
direction perpendicular to the basal plane, that is, $\rho ' <
\rho$, where $\rho$ and $\rho '$ are the tetrahedron side lengths
within and out of the basal plane, respectively, see
Fig.~\ref{fig:lattice}. The fact that the 3D nearest-neighbor
interaction can be made isotropic along the bonds by such slight
engineering of the tetrahedra is crucial, and so the pyrochlore
lattice is the geometry of choice.

We now consider a simple model of dipolar
bosons confined in deep optical lattices in the Hubbard limit.  The
resulting hopping term (tunable through the optical lattice depth)
is nearest-neighbor while the interaction has both on-site and
extended terms \cite{Goral}.  We omit the smaller next-nearest
neighbor interaction terms generated by the long range part of the
dipolar interaction. Working in the insulating limit we may take the
hopping energy gain (in units of the on-site energy) to be much
smaller than the average number of particles per site.  We then
approximate the system by a quantum rotor model:
\begin{equation}
H=-t\sum_{\langle l
l^{\prime}\rangle}b_{l}^{\dagger}b_{l^{\prime}}^{\vphantom{\dagger}}
+ {\rm{h.c.}} + U\sum_{l}n_{l}^2 + V\sum_{\langle
ll^{\prime}\rangle}n_{l}n_{l^{\prime}} \label{Hamiltonian1}.
\end{equation}
Here $\langle l l^{\prime}\rangle$ denotes a nearest-neighbor pair
of sites. $b_{l}^{\dagger}=\exp{(i\theta_{l})}$ is the bosonic
creation operator and $n_l=\partial/ i\partial \theta_l$ counts
the excess number of bosons at a site, $l$.  The dominant
contribution to $U$ arises from the $s$-wave interaction while the
dipolar interaction supplies the largest contribution to $V$. We
assume that both $U$ and $V$ are large, and, in what follows, we
will require $U$ to be slightly larger than $V$. Here and in the
following we omit the renormalized chemical potential term.

{\em Mapping Onto Gauge Theory in $2$D Kagom\'{e} lattice.}
Eq.~\ref{Hamiltonian1} can be
written in a slightly different form more amenable to mapping onto a
gauge theory. We take out a piece, $2U_0$, of the on-site
interaction and group it together with $V$. Defining
$N_{\triangle}\equiv\sum_{l\in\triangle}n_{l}$ and $u\equiv U-2U_0$
, we find, for $V=2U_0$,
\begin{equation}
H=-t\sum_{\langle l
l^{\prime}\rangle}b_{l}^{\dagger}b_{l^{\prime}}^{\vphantom{\dagger}}
+ {\rm{h.c.}} + u\sum_{l}n_{l}^2 +
U_0\sum_{\triangle}N_{\triangle}^{2}. \label{Hamiltonian2}
\end{equation}
Here, $\sum_{\triangle}$ sums over all triangles in the lattice,
and $N_{\triangle}$ counts bosons on a given triangle. In this
form, we call this Hamiltonian a `cluster' Hubbard model where the
fluctuations in the number of bosons on each local cluster
(triangle) costs energy. This property enables a natural map to a
gauge theory which enforces a local constraint.

Eq.~\ref{Hamiltonian2} describes the same Hamiltonian as the one
proposed before \cite{Senthil1, Senthil2} for a microscopic model of
bosons on a square lattice and its 3D generalization, the
corner-sharing octahedra.
We focus here, however, on the
Kagom\'{e} lattice, for which the same Hamiltonian has an exact
implementation in terms of a nearest-neighbor Hubbard model for $U
\sim V$.
The mapping of
Eq.~\ref{Hamiltonian2} on a $(2+1)$-D $U(1)$ lattice gauge theory is
essentially the same as that on a square lattice. The details of the
mapping, however, are different, which we briefly describe below.

A mapping onto gauge theory becomes possible when $U_0$ is the
largest energy scale in the problem.  The boson number on each
triangle is then \textit{locally} conserved. This implies
$N_{\triangle}=0$ on each triangle, meaning that the fluctuation in
the boson number around the mean value set by the chemical potential
vanishes. Working in the limit $U_0\gg t, u$ and doing degenerate
perturbation theory in $t$ and $u$, we find, to first order in $u$
and third order in $t$, the following effective Hamiltonian in the
ground state sector $N_{\Delta}=0$,
\begin{equation}
H_{\rm{eff}}=u\sum_{l}n_{l}^2 -3 \frac{t^3}{U_0^2}
\sum_{\hexagon}b_1^{\dagger}b_2^{\vphantom{\dagger}}b_3^{\dagger}
b_4^{\vphantom{\dagger}}b_5^{\dagger}b_6^{\vphantom{\dagger}}.
\label{Hamiltonian3}
\end{equation}
Here, $\sum_{\hexagon }$ sums over all hexagons inscribed in the
triangles, and $1$ through $6$ indicate the corners of one such
representative hexagon, see Fig.~\ref{fig:lattice}. This effective
Hamiltonian is derived by locally conserving the number of bosons on
each triangle. If a boson hops from site $6$ to site $5$, then
another boson must hop from site $4$ to site $3$ to conserve the
number of bosons on the triangle with one side defined by the bond
$(4-5)$, so that $N_{\Delta}$ remains zero on that triangle. In
turn, as is clear from Fig.~\ref{fig:lattice}, a third boson must
hop from site $2$ to site $1$ to preserve the same constraint.
Finally, since the $u$-term of Eq.~\ref{Hamiltonian2} does not
fluctuate $N_{\Delta}$, it comes in linear order in
Eq.~\ref{Hamiltonian3}.

To see that this Hamiltonian exactly maps onto a lattice gauge
theory on the dual lattice, we consider the dual lattice defined by
the centers of the triangles. This is a hexagonal lattice whose
sites we denote by $r, r^{\prime}$ etc. The original Kagom\'{e}
sites now fall on the \textit{links} of the dual lattice. We take
the representative hexagon, $a$ through $f$, as shown in the left
panel of Fig.~\ref{fig:lattice}, and identify the bosons on
Kagom\'{e} site $1$ with the link $(ab)$ etc.

On the dual lattice, by a change of notation, $H_{\rm{eff}}$
reads,
\begin{equation}
 H_{\rm{eff}}=u\sum_{\langle r r^{\prime}\rangle}n_{rr^{\prime}}^2
 -3 \frac{t^3}{U_0^2}\sum_{\hexagon }b_{ab}^{\dagger}b_{bc}^{\vphantom{\dagger}}
 b_{cd}^{\dagger}b_{de}^{\vphantom{\dagger}}b_{ef}^{\dagger}
 b_{fa}^{\vphantom{\dagger}}.
 \label{Hamiltonian4}
 \end{equation}
Next, noticing that the hexagonal lattice is bipartite, we define a
field $a_{rr^{\prime}}=\theta_{rr^{\prime}}$ if $r\in A$ and
$r^{\prime}\in B$ and $a_{rr^{\prime}}=-\theta_{rr^{\prime}}$ if
$r\in B$ and $r^{\prime}\in A$, where $A$ and $B$ are the two
interpenetrating sublattices, and
$b_{rr^{\prime}}=\exp{({i\theta_{rr^{\prime}}})}$ as discussed after
Eq.~\ref{Hamiltonian1}. We also define the conjugate field variable
$e_{rr^{\prime}}=n_{rr^{\prime}}$ if $r\in A$ and $r^{\prime}\in B$
and $e_{rr^{\prime}}=-n_{rr^{\prime}}$ if $r\in B$ and
$r^{\prime}\in A$ ($e$ and $a$ are conjugate fields since $n$ and
$\theta$ are). Finally, elevating $a$ and $e$ to vector fields
$a_{r\alpha}=a_{r,r+\alpha}$, $e_{r\alpha}=e_{r,r+\alpha}$, where
$\alpha$ indicates the nearest neighbor vectors, it is
straightforward to see that $H_{\rm{eff}}$ can be written as the
Hamiltonian for the $(2+1)$-D compact $U(1)$ gauge theory
\cite{Kogut} on the dual lattice,
\begin{equation}
 H_{\rm{eff}}=u\sum_{r \alpha }e_{r\alpha}^2 -6 \frac{t^3}{U_0^2}\sum_{\hexagon}
\cos(\nabla\times \vec{a}),
 \label{Hamiltonian5}
 \end{equation}
with the Gauss's law constraint
$N_r=N_{\triangle}=\sum_{r^{\prime}\in
r}n_{rr^{\prime}}=\eta_r\nabla . \vec{e}$,
and $\eta_r =1$ if $r\in A$ and $\eta_r =-1$ if
$r\in B$. 
We emphasize that Eqs.~(\ref{Hamiltonian4}, \ref{Hamiltonian5}) are
simply rewritten forms of Eq.~\ref{Hamiltonian3} on the dual
lattice.
Recall that a large $U_0$ implied the constraint on $N_{\Delta}$ in
Eq.~\ref{Hamiltonian2} leading to Eq.~\ref{Hamiltonian3}. Hence, for
an optical lattice implementation of the gauge theory, one simply
needs to implement the Hamiltonian in Eq.~\ref{Hamiltonian1} on
appropriate lattices with both $U$ and $V$ large, with $U$ slightly
larger than $V$. The resulting low energy theory is automatically a
gauge theory on the dual lattices, and so allows the exotic Coulomb
phase and other fractionalized phases \cite{Senthil1, Senthil2,
Fradkin, Kogut} in addition to the conventional insulators
\cite{Goral}. In 2D, however, it is well known that the Coulomb
phase is unstable at long length scales
and is smoothly connected to the conventional Mott insulator.

{\em Mapping Onto Gauge Theory in $3$D pyrochlore lattice.}
We start with formally the same Hamiltonian as
Eq.~\ref{Hamiltonian1}, but on a pyrochlore lattice.
From Eq.~\ref{Hamiltonian1}, one straightforwardly gets the
`cluster'-Hubbard model, Eq.~\ref{Hamiltonian2}, with
$N_{\triangle}$ replaced by $N_{\rm{T}}$, the number of bosons on a
tetrahedron. The mapping to gauge theory on the dual diamond
lattice, which is bipartite, but not Bravais, is identical and one
ends up with Eq.~\ref{Hamiltonian5} as the final effective theory.
To have a description of the dual
lattice in terms of a Bravais lattice, we consider below only one
FCC sublattice of the diamond lattice (centers of only one class of
tetrahedra on the pyrochlore). Note that by counting the centers of
only one class, and allowing an index $\alpha$ to indicate all four
corners of a tetrahedron from its center, we can count all the
pyrochlore sites (diamond links).

In $(3+1)$-D, the Coulomb phase, which is an insulator, is described
by the Gaussian expansion of the cosine term of the gauge theory
Hamiltonian, and is stable for $t\gg \sqrt[3]{uU_0^2}$, since the
topological defects -- the monopole configurations of the gauge
field -- are suppressed in the Coulomb phase. This is analogous to
the gaussian expansion of the Hamiltonian of the $XY$ model in the
Kosterlitz-Thouless phase where the topological defects -- the
vortices -- are suppressed.
 Notice that a Gaussian
expansion of the cosine term yields a Hamiltonian formally the same
as that in classical electrodynamics, with an associated gapless
`photon' mode. However, this is a gapless mode of the gauge field of
the bosonic field operators, and {\textit{not}} of an external
electromagnetic field. This mode, which \textit{emerges} in the low
energy theory, should therefore be observable in the long-wavelength
boson density-density correlation functions, and hence in noise
correlations, given by the correlation functions of the $e$-fields
(Recall that $e_{rr^{\prime}}$ is simply related to
$n_{rr^{\prime}}$). This mode distinguishes itself from the
conventional phonon mode resulting from spontaneously broken
symmetry in strongly correlated, bosonic models because it appears
in the {\em absence} of spontaneous symmetry breaking and is
therefore an emergent phenomenon. Furthermore, this mode appears in
a range of parameters ($U_0\gg t \gg \sqrt[3]{uU_0^2}$) intermediate
between that for a superfluid ($t \gg U_0, u$) and the conventional
insulators ($U_0, u \gg t$).

{\em Noise Correlations in the Coulomb Phase.} We now consider the
long wavelength behavior of noise correlations in terms of the
$e$-field correlation functions.
\begin{figure}
\includegraphics[clip,width=3in]{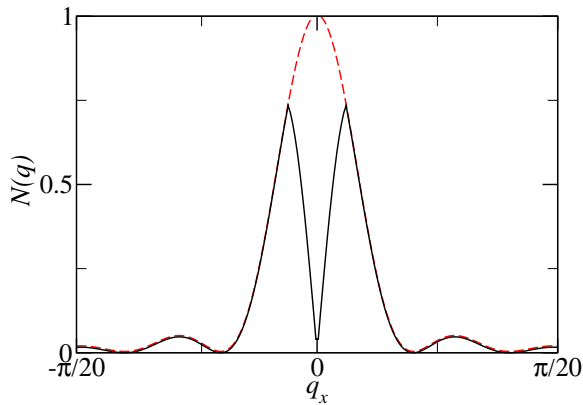}
\caption{Expected behavior of the normalized noise correlation
function with $q_x$ in the Mott insulating (dashed curve) and
Coulomb (solid curve) phases around the first Brillouin-zone center
of the FCC lattice. The core of the Mott insulating peak is
suppressed due to the $c|q|$ non-analiticity provided by the gapless
`photon' at long wavelengths in the Coulomb phase. The behavior, at
large $q$, is expected to cross over to that of the Mott insulator.
The plot is for a finite sized system with 100 sites along each
direction with a small $q$ slope $c=40$ (with the appropriate form
factor set to unity) and an arbitrarily chosen cut-off at
$q_x=3\pi/500$. } \label{fig:noise}
\end{figure}
To compute the correlation function of the $e$-fields, we work in
the Coulomb gauge where $a_{\tau}=0 $ and $\nabla \cdot \vec{a}=0$.
We then expand the cosine in the second term of
Eq.~\ref{Hamiltonian5} to quadratic order and use $e=\frac{\partial
a}{\partial \tau}$. Working in the continuum limit, the $a$-field
correlator at momentum $k$ and (imaginary) frequency $\omega$ is
given by,
\begin{equation}
\langle
a_i(k,\omega)a_j(-k,-\omega)\rangle\sim\frac{\delta_{ij}-k_ik_j/k^2}{\omega^2+c^2k^2},
\label{a-correlator}
\end{equation}
where $c^2=(6 t^3)/(uU_0^2)$.  This reveals the mass-less
emergent `photon' mode. The equal-time
correlator $\langle
e_{r\alpha}e_{r^{\prime}\alpha^{\prime}}\rangle$, where $\alpha,
\alpha^{\prime}$ are unit vectors for a tetrahedron, can be found
from here. This gives the boson density-density correlation
function. We find,
\begin{eqnarray}\langle
n_{r\alpha}n_{r^{\prime}\alpha^{\prime}}\rangle &=& \langle
e_{r\alpha}e_{r^{\prime}\alpha^{\prime}}\rangle
=\alpha_i\alpha^{\prime}_j\langle
e_{ri}e_{r^{\prime}j}\rangle\nonumber\\ &\sim& \alpha_i
\alpha^{\prime}_j\sum_{k}c|\vec{k}|(\delta_{ij}-\frac{k_ik_j}{k^2})e^{i\vec{k}\cdot(\vec{r}-\vec{r}^{\prime})}.
\label{e-correlator1}
\end{eqnarray}
A straightforward evaluation of this integral in the continuum gives
the special angular structure of the Coulomb phase density
correlation function at large distances \cite{Balents}. However, as
described below, for noise correlation, one is interested in the
function in momentum space itself.

In time of flight imaging of an insulating state of spinless particles,
averaging the noise between shot-to-shot images of
the particle distribution released from
optical lattices reveals a quantity proportional to the following
second order correlation function\cite{Eugene}:
\begin{equation}
N (\vec{q})\sim \sum_{r\alpha
r^{\prime}\alpha^{\prime}}e^{i\vec{q}\cdot(\vec{r}+\vec{\alpha}-\vec{r}^{\prime}-\vec{\alpha}^{\prime})}\langle
n_{r\alpha}n_{r^{\prime}\alpha^{\prime}}\rangle,
\end{equation}
where we, for simplicity, omit the $\vec{q}=0$ delta function due to
normal ordering. In a Mott insulator, which should be compared with
the Coulomb insulator, $\langle
n_{r\alpha}n_{r^{\prime}\alpha^{\prime}}\rangle$ is a constant, and
so, for an infinite system $N (\vec{q})\sim \sum_n f(\vec{q})
\delta(\vec{q}-\vec{q}_n)$ where $\vec{q}_n$'s are the reciprocal
lattice vectors of the FCC lattice, and $f(\vec{q})$ is the form
factor of the tetrahedron basis, $f(\vec{q})=\sum_{\alpha
\alpha^{\prime}}\exp{[i\vec{q}\cdot(\vec{\alpha}-\vec{\alpha}^{\prime})]}$.
This produces $\delta$-function peaks at the reciprocal lattice
vectors which are the centers of the Brillouin zones. For the
Coulomb phase, we know the behavior of $\langle
n_{r\alpha}n_{r^{\prime}\alpha^{\prime}}\rangle$ only for large
$|\vec{r}-\vec{r}^{\prime}|$. At smaller length scales, the
calculation is quite involved since the full tetrahedron lattice
correlators have to be used \cite{future_pub}. The long distance
behavior, however, is sufficient to give the leading singularities
at the Brillouin zone centers. From the integrand of
Eq.~\ref{e-correlator1} we see that, close to the zone centers, the
leading singularity is a cusp, $N(\vec{q})\sim c|\vec{q}|$, see
Fig.~\ref{fig:noise}.

In conclusion, we have shown that compact $U(1)$ lattice
gauge theories can be simulated in suitably designed cold atom
optical lattices using dipolar bosons. Exciting exotic phenomena,
the Coulomb phase with emergence of artificial photons being an
example, could be studied experimentally following our suggestion.

We thank L. Balents, H. Buchler, C. Zhang, A. Vishwanath and O.
Motrunich for discussions.
This research is supported in part by the National Science
Foundation under Grant No. PHY99-07949, and ARO-DTO, ARO-LPS, and
NSF. TS was very generously supported through a DAE-SRC Outstanding
Investigator Award in India.


\end{document}